\documentclass[journal,twocolumn]{IEEEtran}
\usepackage{epsfig,makeidx,color}
\usepackage{amsmath,amssymb,bbm}
\usepackage{cite,graphicx,lipsum,subfigure}
\usepackage{enumerate}
\usepackage[switch,pagewise]{lineno}
\usepackage{hyperref}
\hypersetup{
        colorlinks = true,
        citecolor=blue,
}
\def\be{ \begin{equation} }
\def\ee{ \end{equation} }
\def\bea{ \begin{eqnarray} }
\def\eea{ \end{eqnarray} }

\def\b0{{\bf 0}}

\ifCLASSOPTIONonecolumn
\else

\fi

\begin{document}

\title{Grant-Free Random Access in Machine-Type
Communication: Approaches and Challenges}

\author{Jinho Choi, Jie Ding, Ngoc
Phuc Le, and Zhiguo Ding}


\maketitle
\begin{abstract}
Massive machine-type communication (MTC) is expected to 
play a key role in supporting
Internet of Things (IoT) applications such as smart cities, 
smart factory, and connected vehicles through cellular networks.
MTC is characterized by a large number of MTC
devices and their sparse activities, 
which are 
difficult to be supported by conventional approaches 
and motivate the design of new access technologies.
In particular, in the 5th generation (5G), grant-free
or 2-step random access schemes are introduced 
for MTC to be more efficient by reducing signaling overhead.
In this article, we first introduce grant-free
random access and discuss how it can be modified
with massive multiple-input multiple-output (MIMO) 
to exploit a high spatial multiplexing gain.
We then explain preamble designs
that can improve the performance
and variations based on 
the notion of non-orthogonal multiple access (NOMA).
Finally, design challenges of grant-free random access
towards next generation cellular systems
are presented.
\end{abstract}


\ifCLASSOPTIONonecolumn
\baselineskip 24pt
\fi

{\IEEEkeywords
Machine-Type Communication,
Random Access, Multiple-Input Multiple-Output} 

\section{Introduction}

In order to support the connectivity of
a large number of devices and sensors (or things)
within cellular systems, 
machine-type communication (MTC)
has been actively studied.
Thanks to MTC for massive
access \cite{Ali17}, various applications of 
the Internet-of-Things (IoT) with  myriads of devices 
deployed over a large geographic area 
(e.g., smart factory,
smart cities, and intelligent transportation) become possible.

MTC differs from human-type communication (HTC)
in a number of ways. For example, MTC 
is expected to support
massive and 
burst traffic due to a massive number of devices and sparse 
activity of each device. In particular,
MTC is expected to support up to 1 million devices per km$^2$,
and devices usually have low duty cycle and transmit small payloads.
Clearly, HTC's transmission schemes not suitable for 
MTC as they result in a high signaling overhead. 
Therefore, it is expected that media access protocols 
based on random access 
will be preferred to provide MTC services. 
In addition, 
MTC should be very efficient 
in providing connectivity with limited bandwidth, since
there are a large number of MTC devices and sensors.
As such, there is increasing interest in the 
development of various new transmission methods 
and media access protocols with low signaling overhead 
and high spectral efficiency for MTC.

Random access channel (RACH) procedure is widely 
used in cellular systems to establish a connection from 
a user to its base station (BS) and to be assigned channel 
resources for uplink transmission by the user. 
RACH procedure is also used for MTC. 
When a device becomes active 
and wants to transmit its data, it can choose
one of pre-defined preambles and transmit it, which
is the first step of RACH procedure. 
Since multiple devices can be active at the same
time and choose any preambles randomly,
RACH procedure can be seen as a contention-based
access. In particular, preamble
collision happens when multiple active
device chooses the same preamble. Due to 
multiple preambles, RACH procedure can
be modeled as a multichannel 
(slotted) ALOHA, where each preamble
can be seen as a channel.
The performance of MTC is limited due to preamble
collision and the probability of preamble collision grows 
exponentially with the number of active devices.
Thus, access control and dynamic
resource allocation become crucial in MTC
\cite{Ali17}.

Recently, a new random access scheme,
called 2-step random access, is proposed for MTC 
in the 5th generation (5G) by
the 3rd generation partnership project (3GPP) to 
further reduce signaling overhead \cite{3GPP_MTC_18} \cite{Kim20}. 
In 2-step random access, unlike RACH procedure,
an active device
does not wait for a response from the BS after 
transmitting the preamble, and immediately transmits data packet.
Thus, it is expected to shorten access delay, which
is desirable for low latency applications,
and improve the throughput so that more devices
can be supported for massive MTC.
2-step random access is also called
grant-free (GF) random access, because
active devices transmit data packets without
obtaining any reserved channel resources. 

Multiple-input multiple-output (MIMO) has been extensively 
investigated to improve the spectral efficiency
of wireless communications over the past 30 years. 
In particular, the notion of massive MIMO
was exploited
to take advantage of the high spatial multiplexing gain 
of BSs equipped with large-size antenna arrays
\cite{Lu_mMIMO}. Thus,
massive MIMO can also be a solution to support
a large number of devices in MTC
with limited bandwidth together with GF
random access \cite{deC17} \cite{Senel18}.
A salient feature of 
GF or 2-step random access 
facilitated by massive
MIMO is that both the two steps 
can be carried out on a single channel resource,
which can further make 
GF random access simple and efficient.

Like 4-step random access, an
active device is to transmit a randomly 
selected preamble in 2-step or GF random access,
which also allows for BSs to estimate device's channel
state information (CSI) or channel vector
for successful decoding of data packets in massive MIMO.
Consequently, preamble design becomes crucial 
not only for random access, but also for 
successful decoding
in GF random access with massive MIMO.
More preamble sequences are needed to support more devices
in GF random access, 
which may require more radio resources.
Thus, efficient approaches to generate more preambles
are highly desirable. In addition, 
non-orthogonal multiple access (NOMA)
\cite{Choi_JSAC} can be employed 
for 
the spectral efficient GF random access
to improve in spectral efficiency.
In this article, we 
present the key ideas of GF random access 
with improved performance by massive MIMO,
semi-GF transmission facilitated by NOMA, and
discuss preamble designs and
the application of NOMA for variations of GF random access.

Note that a comprehensive overview
of 2-step random access standardized for MTC in 5G
\cite{3GPP_MTC_18} can be found in \cite{Kim20}.
As such,
in this article, we mainly focus on
fundamentals of GF
or 2-step random access with massive MIMO
and discuss possible its variations.
In addition, design challenges of GF random access 
are discussed from
the perspective of the 6th generation 
(6G) \cite{Dang20}.


In the remainder of this article, we first explain
4-step and 2-step random access schemes for MTC.
We then show how massive MIMO can be employed for 2-step
random access to significantly improve spectral efficiency.
Since the performance of 2-step random access
also depends on preambles, we discuss preamble designs.
A number of variations of 2-step random access are also discussed.
We also identify challenges of GF random access
towards next 
generation of mobile networks, i.e., 6G.

\section{Random Access Protocols in MTC}

\begin{figure}
\begin{center}
\includegraphics[width=0.4\textwidth]{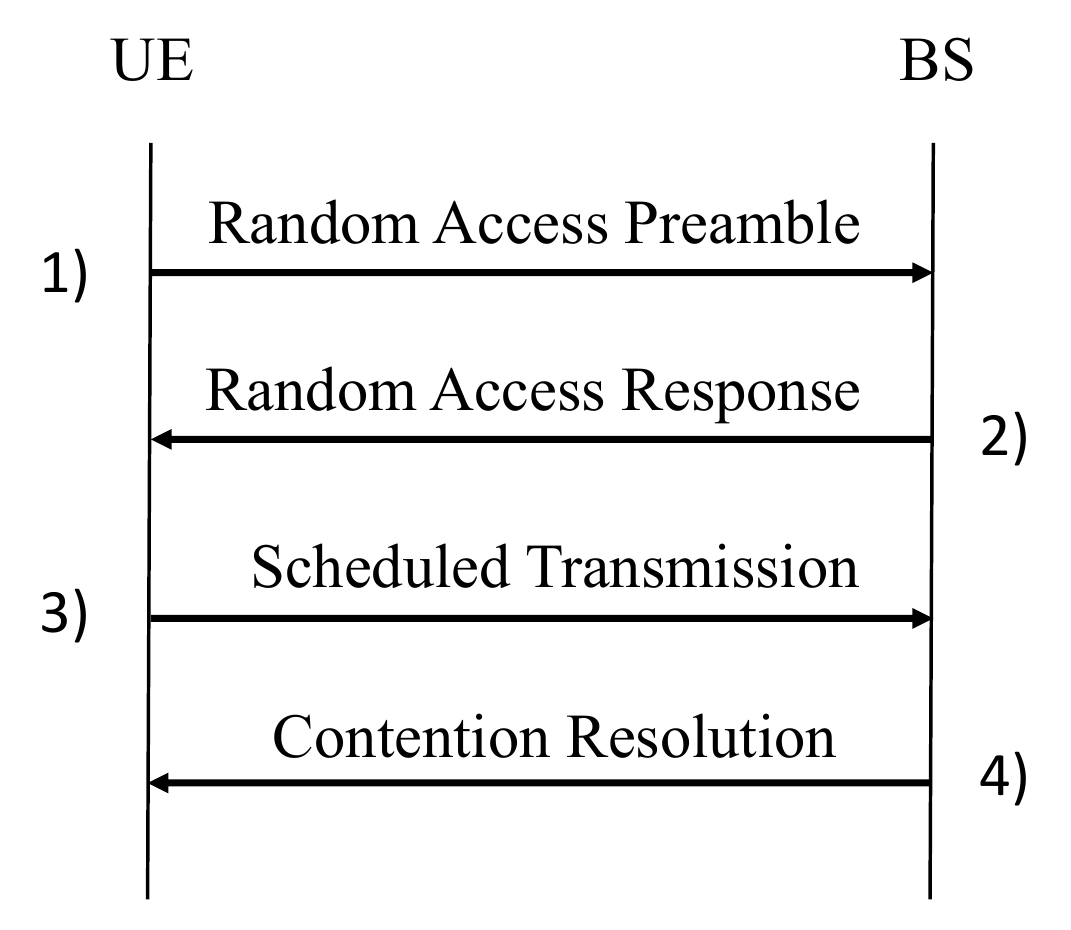}
\end{center}
\caption{4-step random access procedure in MTC.}
                \label{Fig:4step}
\end{figure}

In MTC, uplink transmissions can be carried out
based on 4-step random access procedure,
which is illustrated in Fig.~\ref{Fig:4step}.
The first step is random access
to establish connection to the BS
with a pool of preambles consisting of $L$ sequences.
In the first step, an active user equipment (UE)
transmits a randomly selected a preamble
on physical random access channel (PRACH).
In the second step, the BS detects the preambles 
transmitted by active UEs and sends responses. Once
an active UE is connected to the BS, it can transmit 
data packets
in the third step
on dedicated resource blocks (RBs)
or channels\footnote{We will use
the terms resource block and channel interchangeably.},
which are physical uplink shared channel (PUSCH).

Compared with 4-step random access approaches,
2-step random access \cite{3GPP_MTC_18}, which is also referred
to as GF random access, can be more efficient
thanks to low signaling overhead when 
UEs or devices have short messages to transmit.
In 2-step random access, an active device
does not wait for a response from the BS, 
i.e., Step 3 is removed and
only Steps 1 and 4 in Fig.~\ref{Fig:4step}
are used, where Steps 1 and 3
are combined into Step 1.
That is, in Step 1, an active
device is to transmit a preamble 
on PRACH and then transmit a payload
on PUSCH in a time division multiplexing 
(TDM) manner. 

It is also possible to reserve certain
preambles for UEs that require contention-free random access.
In this case, the BS sends 
the information of reserved preambles
and PUSCH assignment prior to any
random access procedure to avoid collision.

\section{Massive MIMO for GF Random Access}

In massive MIMO, thanks to a
large number of service antennas co-located at a 
BS, the channel hardening (i.e., the effect of small-scale fading 
is averaged out and devices' channels behave 
deterministic like wired channel) and favorable 
propagation (i.e., the propagation
channels to different devices 
become orthogonal, which makes different devices
distinguishable in space)
can be exploited \cite{Lu_mMIMO}. 
Taking advantage of these two features, 
it is possible to 
spatially separate signals 
that are simultaneously transmitted
by 
multiple MTC devices that share the same channel
in GF random access. 

\begin{figure}[thb]
\begin{center}
\includegraphics[width=0.4\textwidth]{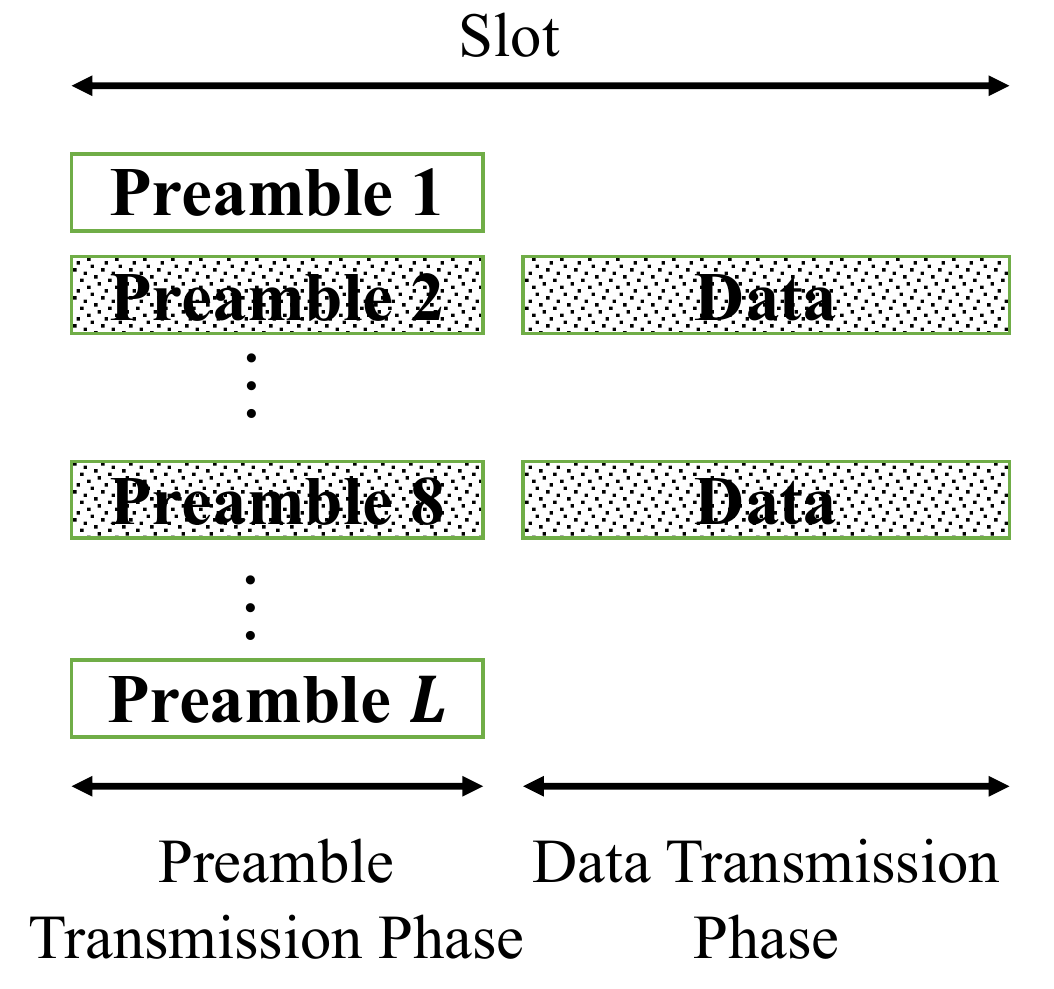}
\end{center}
\caption{A slot consisting of two phases, namely
preamble and data transmission
phases, for a 2-step random access scheme (the shaded
blocks represent transmitted signals, i.e., there are
two active devices transmitting preambles 2 and 8).}
        \label{Fig:two_phase}
\end{figure}

Massive MIMO can make
GF random access
further efficient by allowing transmissions of
preambles and payloads on the same channel resource.
In particular,
for the first step that has
preamble and data transmission phases,
one time slot (or one channel resource) can be used
as illustrated in Fig~\ref{Fig:two_phase}.
Each active device chooses
a preamble from a pool of $L$ preambles uniformly at random,
and transmit it in the preamble transmission phase.
In the data transmission phase, a data packet is then transmitted.
Since all the devices are synchronized in MTC,
it is expected that the length of data packet is the same
for all devices (or the length of data transmission phase
is decided by the maximum length of data packet among
all the devices). 
Because of the orthogonality among those spatial channels,
the BS is able to decode each of co-existing signals
transmitted by multiple active devices on the same
channel resource
during the data transmission
phase without
any orthogonal multiple access
schemes (e.g., time division multiple access)
\cite{deC17} \cite{Senel18}. 

To decode co-existing signals via beamforming, 
the BS needs to
estimate the spatial channels of active devices. 
An active device's
channel can be estimated
if its preamble is not transmitted by other
active devices (i.e., there is no
preamble collision). As a result, 
the performance of GF random access
with massive MIMO hinges on how to minimize 
preamble collisions.


\section{Preamble Designs}

Although massive MIMO
can offer a high spatial multiplexing gain,
the number of active devices that can successfully
transmit their payloads is limited by
the number of preambles, $L$,
in GF random access. 
Thus, to fully exploit
a high spatial multiplexing gain,
it is desirable to 
generate a larger number of preambles
to support a large number of devices without increasing
the length of preamble transmission phase.
In this section, we 
present a few approaches that can effectively
generate more preambles.

\subsection{Multiple Preambles}

To mitigate preamble collision 
with a finite number of orthogonal preambles,
preamble designs
can be based on multi-preamble transmissions
within a transmission frame \cite{deC17}, 
where a frame consists of multiple slots
and a slot is further divided into two sub-slots
for preamble and data transmission phases as in Fig.~\ref{Fig:two_phase}.
Each device has a unique preamble-hopping pattern across 
multiple slots, which is assigned by the BS. 
Let $B$ denote the number of slots per frame.
Provided that there are $L$ orthogonal preambles,
there can be $L^B$ different preamble-hopping patterns.
In other words, up to $L^B$ devices can be supported
with different preamble-hopping patterns.


A similar multi-slot transmission structure can 
be found in \cite{Coded_pilot}, where 
the notion of coded random access is employed.
In coded random access,
each active device transmits multiple randomly selected 
preambles within a frame, while the same
data packet is repeatedly transmitted
(which differs from the approach in \cite{deC17}
where a device transmits different data packet in each slot).
Within a frame, there can be 
time slots with collision-free transmissions
that allow interference-free channel
estimation of some devices and then successful decoding of their
data packets. The BS then applies
successive interference cancellation
(SIC) in order to remove 
the signals of these devices in all the slots.
After SIC, the BS can further find 
slots with collision-free transmissions and perform
channel estimation and decoding, and so on.
The same process is to be repeated until the BS cannot
find any slot with collision-free transmission.

With a finite number of orthogonal preambles,
these multi-slot transmission schemes have 
the potential 
to increase the number of active devices by
mitigating/resolving the preamble and data collision 
in GF random access
at the cost of power consumption at devices due to multiple 
transmissions.

\subsection{Non-Orthogonal Preambles}

A naive approach to preamble design to increase
the number of preambles without increasing their length
is based on non-orthogonal sequences.
For example, if Zadoff-Chu sequences are used,
the number of preambles can be $L (L-1)$
with a cross-correlation of $\frac{1}{\sqrt{L}}$.
Gaussian random sequences can also be
used as non-orthogonal preambles,
which can allow each device in the cell to have 
a unique non-orthogonal preamble as its signature identification, 
because the number of sequences can be arbitrarily large.

While making use of non-orthogonal sequences for preambles
can increase the number of preambles to lower
the probability of preamble collision,
there is a problem in obtaining a good
channel estimate that is essential in massive MIMO.
Due to non-zero cross-correlation, the channel
estimate is contaminated, which leads to 
degraded decoding performance.
As a result,
it is demonstrated in \cite{Ding20b} that 
non-orthogonal 
preamble does not necessarily provide 
a higher success probability than its orthogonal counterpart. 

It is also noteworthy that 
compared to orthogonal preambles, 
the detection of non-orthogonal preambles 
requires a high computational complexity
and more advanced signal processing techniques \cite{Senel18}. 

In \cite{Choi_TWC20}, in order to take 
both advantage of non-orthogonal and orthogonal preambles, 
superpositioned preambles (S-preambles) are studied.
Each device uses a linear combination 
of few different orthogonal preambles to generate an S-preamble. 
For example, superposition of 2 orthogonal preambles
can be considered, which can generate 
$\frac{L(L-1)}{2}$ S-preambles.
In general, S-preambles can be viewed as structured
non-orthogonal preambles.
Note that unlike the approach based on multi-preamble transmissions
in \cite{deC17}, each active device transmits
one S-preamble. 
Therefore, the length of the preamble transmission phase in a slot
does not increase.
Due to the feature of S-preambles,
it is possible to use low-complexity algorithms for the channel estimation 
(as the case of orthogonal preambles)
with lowering 
the preamble collision probability (as the case of non-orthogonal preambles). 
Consequently, a better performance can be achieved without
increasing the complexity of channel estimation.



\section{Semi-GF Random Access}


Semi-GF transmission can offer more refined 
admission control compared to GF schemes \cite{8662677}, 
and yield less system overhead compared to grant-based schemes. 
In particular, in cellular systems, 
BSs can play a crucial role in providing a certain
level of user coordination to improve
the performance. In this section, we discuss various 
semi-GF approaches 
that can provide high spectral efficiency than pure
GF approaches.

\subsection{Semi-GF Transmission based on NOMA}

The use of semi-GF transmission is motivated by the following 
two facts. Firstly, as mentioned earlier, 
in cellular systems, it is possible to take advantage of
user coordination by BSs instead of  
relying completely on pure random access, 
such as GF transmission. Secondly, 
semi-GF transmission is able to seek more cooperation 
among grant-based and GF users
so that limited radio resources can be 
more efficiently utilized.
Based on these motivations,
semi-GF transmission 
can be considered in order to encourage 
the spectrum cooperation among the grant-based 
and GF users, where the BSs are 
employed as an essential component for admitting GF users. 

\begin{figure}[!t]\vspace{-1em}
\begin{center} \subfigure[ The case with a strong grant-based user   scheduled ]{\label{fig 0 ax}\includegraphics[width=0.45\textwidth]{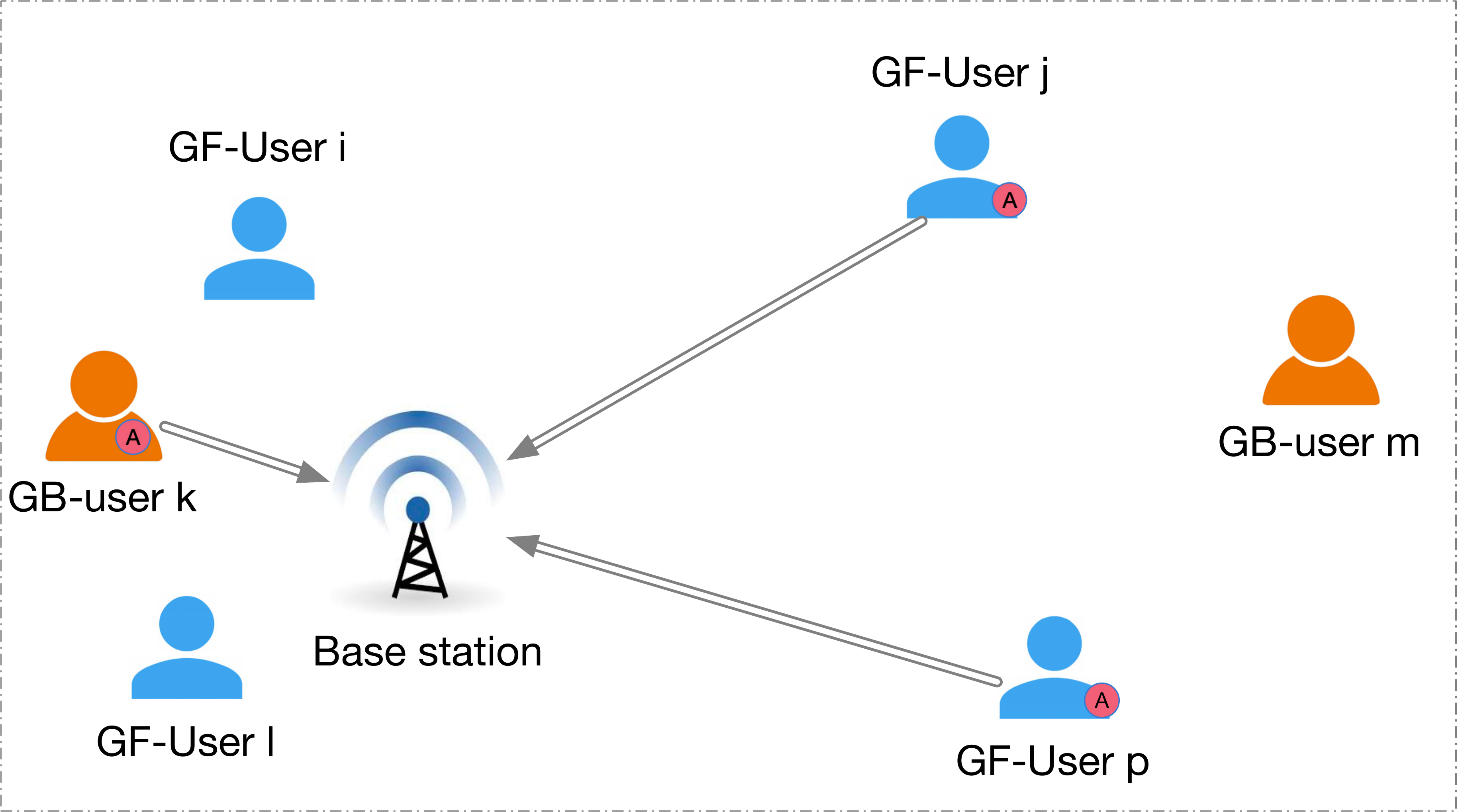}}
\subfigure[The case with a weak grant-based user    scheduled ]{\label{fig 0 bx}\includegraphics[width=0.45\textwidth]{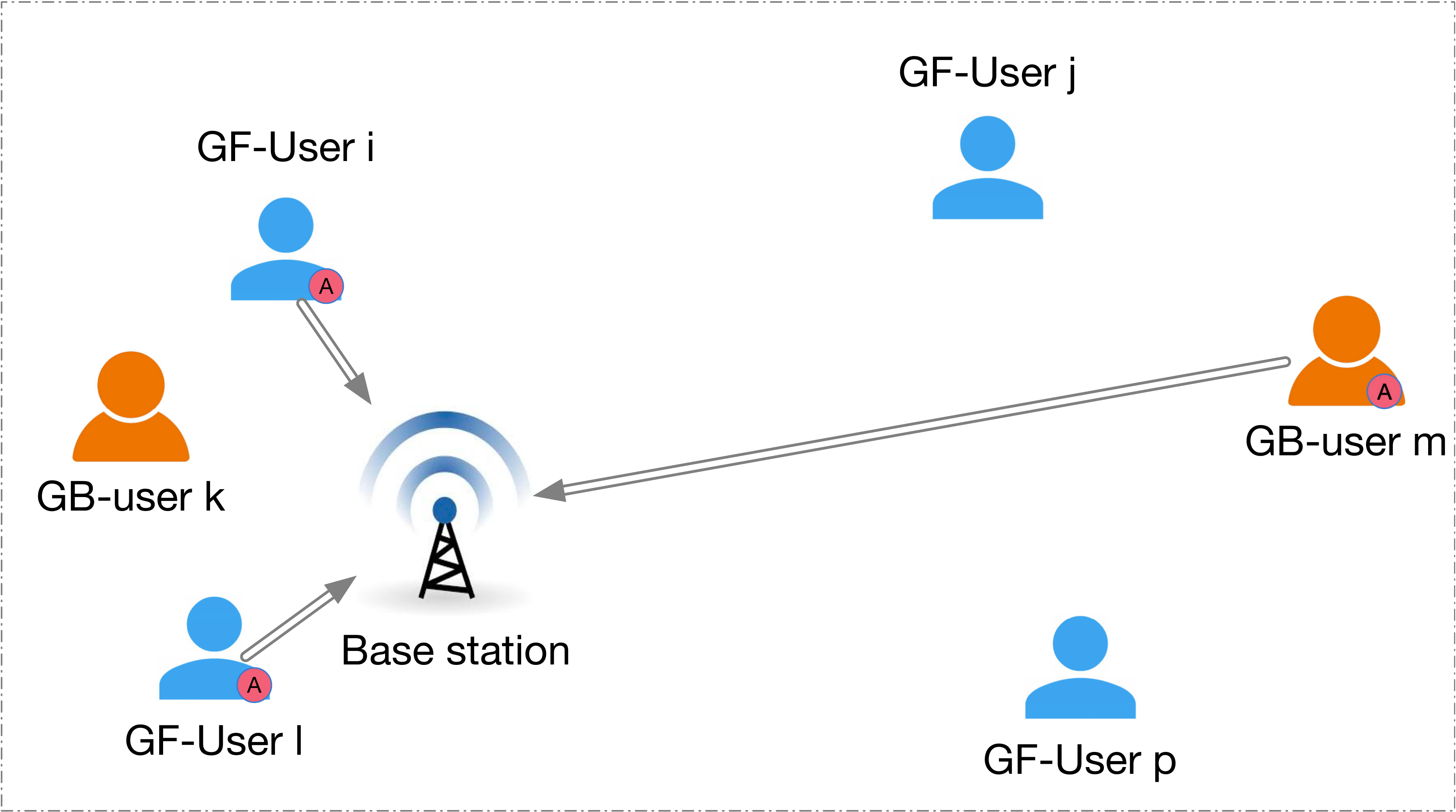}} \vspace{-1em}
\end{center}
\caption{  Illustration for the principle of semi-GF transmission. For the case with a strong grant-based user scheduled, the BS   invites GF users with weak channel conditions to participate in contention.   For the case with a weak grant-based user scheduled, the BS  invites GF users with strong channel conditions to participate in contention.    }\label{fig0}\vspace{-2em}
\end{figure}

Fig. \ref{fig0} describes two extreme cases which 
illustrate an approach of semi-GF transmission
based on NOMA.
In particular, in Fig. \ref{fig 0 ax}, 
for a case in which a grant-based user 
with strong channel gains is scheduled to transmit, 
the BS can broadcast a threshold based on the grant-based user's 
channel information, where only those GF 
users which have channel gains weaker than this threshold  
are allowed to compete and transmit. The BS is to 
decode the grant-based user's signal first
prior to decoding the GF users' signals. 
Clearly, the BS needs to reduce the threshold in order to ensure that 
the grant-based user's experience is not degraded. 
Fig. \ref{fig 0 bx} illustrates another extreme case, 
where a grant-based user with weak channel conditions is scheduled. 
Again the BS broadcasts a threshold. From this, only 
the GF users with the channel gains that are
higher than this threshold can be allowed
to participate in contention. The reason for this 
is that the BS will decode the GF users' signals first 
before decoding the grant-based user's signal. 
Again by changing the threshold, the BS 
is able to adaptively adjust  
the number of admitted GF users in a low system overhead manner.

\subsection{Other Semi-GF Random Access Schemes}

The design of preambles is crucial 
to the practical implementation of semi-GF transmission. 
In \cite{QZhang20}, a semi-GF random access 
with orthogonal preambles was investigated,  
where the BS sends the feedback after receiving preambles
so that active devices with preamble collision do 
not transmit their data packets. This can result 
in a high signal-to-interference-plus-noise ratio (SINR) 
of the active devices that do not experience preamble collision 
and the increase of throughput.
In \cite{Choi_IoT20}, a compressive semi-GF 
random access with non-orthogonal preambles was proposed. 
In this scheme, compressive sensing techniques were used 
to detect non-orthogonal preambles that sent by devices and 
the information including the indices of detected transmitted 
preambles are feedback so that each device can be allocated a unique slot resource to transmit data packet.  
To achieve a higher spectral efficiency, the concept
of NOMA-based GF random access was recently proposed \cite{Ding20_TNOMA}. Unlike conventional GF random access where preambles are time-multiplexed with data, i.e., preamble and data transmissions do not co-exist, NOMA-based GF random access exploits the favorable propagation of massive MIMO to allow co-existing preamble and data transmissions by two different groups of devices in each time slot; such that  the resource utilization efficiency for both preamble and data transmissions can be significantly increased, which improves the access performance for MTC devices.

\section{Challenges towards 6G}

The number of connected devices is expected to exceed 500 billion by 2030,
while the human population
is predicted to be 8.5 billion by the United Nations (UN).
Clearly, the number of devices will be
about 60 times the human population by 2030.
Thus, MTC will play a more important role in 6G and
new MTC schemes are to be developed for a variety of IoT applications using
cell-free architecture, Terahertz communications, three-dimensional networking, and artificial intelligence \cite{Dang20}.
In what follows, we will discuss potential research directions 
and design challenges 
of GF random access in 6G.

\subsection{GF Random Access in Cell-Free Massive MIMO}

Cell-free massive MIMO is regarded as 
a potential evolutionary technology for existing massive MIMO for 6G.
In cell-free massive MIMO, 
the service antennas, in the form of simple 
and low-cost access points (APs), 
are distributed over a wide area without the 
conventional notion of cell boundaries. 
Several studies have shown that 
cell-free massive MIMO is more robust to 
shadow fading effect and hence can provide 
a better coverage than massive MIMO. 

In 6G, GF random access
is to be deployed with cell-free massive MIMO.
However, most of the existing GF random access approaches 
with massive MIMO 
rely on the features of favorable propagation 
and channel hardening, 
which may not be applicable in 
cell-free massive MIMO. 
In particular, different from massive MIMO where the
signals received at all the antennas from a device experience 
identical large-scale fading
due to the fact that all the antennas are co-located, 
the signals at different 
antennas undergo different 
large-scale fading 
in cell-free massive MIMO. 
As a result, the channel hardening 
effect is relatively insignificant. 
On the other hand, cell-free massive MIMO provides 
other unique phenomena such as spatial sparsity 
and macro diversity, which are not available 
in massive MIMO. 
For example, only few antennas sufficiently close to a device
can receive strong signals from the device, 
while the other antennas do not
receive sufficiently strong signals.
This spatial sparsity 
in cell-free massive MIMO needs to be exploited to 
design GF random access 
to support a large number of devices with their
unique spatial sparsity patterns.

Another challenge is to overcome the asynchronous 
reception in cell-free massive MIMO for
GF random access.
Specifically, due to the distributed nature of the network 
and the differences in propagation delays, 
signals received from different devices at 
distributed antennas may not be synchronized
with sufficient precision. 
The asynchronous reception effect may have 
a huge impact on the preamble detection and channel estimation, 
crippling the performance of GF random access. 
To address this issue, 
the preamble design that is robust to 
asynchronous reception 
can be studied together with
tailored phase (or timing) estimation and 
compensation schemes 
to minimize the impact 
of asynchronous reception on the performance
of GF random access.

\subsection{GF Random Access for Massive URLLC}

Ultra-reliable low-latency communication (URLLC) 
has been considered for delay-sensitive  real-time  applications
in 5G with stringent requirements in terms
of reliability and latency. 
In 6G, tighter requirements are expected. 
For example, the target error rate of $10^{-5}$ in 
5G will be even lower to $10^{-7}$ in 6G.
To improve reliability in 6G, 
various diversity techniques including
multi-connectivity schemes
can be employed. 

In 6G, air latency is
expected to be less than $100 \mu$s,
which makes the
random access design 
quite challenging. 
To guarantee a targeted
latency, contention-free random access
can be used with reserved 
preambles in GF random access.
That is, for
a device with a reserved preamble,
there is no re-transmission due to
preamble collision and 
successful transmission of payload can be guaranteed within a target transmission delay time.
Unfortunately, if the number of devices increases in delay-sensitive real-time applications, it can lead to a shortage of reserved preambles. 
Addressing this issue for massive URLLC is more challenging than 
massive MTC
as it also needs to take the extreme reliability 
and latency requirements into account. 
To achieve a scalability-reliability-latency tradeoff 
for massive URLLC, the channel inference 
and traffic prediction can be employed 
to underpin the preamble designs or management, 
where the temporal correlation of wireless 
channels and history knowledge of devices' channel 
and traffic states can be exploited 
to improve the preamble utilization efficiency, 
mitigate the interference originating from preamble collisions, 
and enhance channel estimation performance.

\subsection{GF Random Access for 3D Networks}

6G will include non-terrestrial networks (NTN)
to support ubiquitous and global connectivity.
In MTC, NTN will play a key role
in providing global connectivity for a massive number of devices. 
As a result, it is necessary to develop 
GF random access schemes that are well-suited to 
three-dimensional (3D) networks comprising NTN and conventional
terrestrial networks.
For NTN, there will be nonterrestrial nodes
including unmanned aerial vehicles (UAVs), 
high altitude platform stations (HAPSs), and satellites.
Due to the high mobility and limited energy of nonterrestrial nodes,
the resulting systems
exhibit different channel characteristics and deployment scenarios, 
which poses new challenges for random access mechanisms. 
In 3D networks, one major challenge is to ensure 
the coexistence of terrestrial and nonterrestrial 
devices in GF random access.
As the UAVs
are deployed  in high altitude, 
the UAV-BS channels are usually dominated by line-of-sight (LoS) links, 
which can be significantly different from those of 
terrestrial devices.
In addition, the deployment of UAVs in 
the altitude dimension also has non-negligible 
impact on the aerial-ground interference. 
Thus, new approaches to support the unprecedented 
aerial-ground GF random access are needed, 
where the channel characteristic disparity 
between terrestrial
devices and nonterrestrial 
nodes, e.g., UAVs,
and the deployment of UAVs in 3D networks should be taken into
account in design.

In addition, in some remote areas where devices are randomly 
deployed  (e.g., in forest for bushfire monitoring) without communication
infrastructure, a swarm of UAVs can be used as
gateways to collect data from randomly active devices, which results in UAV-swarm based GF random access.
Different from conventional terrestrial GF random access where the locations of gateways or BSs are usually fixed on the ground, 
the deployment of UAV swarm 
in the altitude dimension can be flexibly and coordinately adjusted according to service requirements and wireless environment change.
Besides, since UAVs operate under
strict resource constraints such
as limited battery capacity, time and energy efficiency become crucial for UAV swarm.
Therefore, optimal deployment 
of UAV swarm needs 
to be carefully investigated to efficiently 
enable UAV-swarm based GF random access in terms of time and energy.

In 3D networks, 
GF random access schemes are preferred 
due to its spectral efficiency and low access delay, which is essential especially when devices are remotely deployed. 
Efficient GF 
random access designs therefore need to 
take into
account channel characteristics of these networks. 
A crucial task for 
GF random access is to reduce the access delay 
or enhance the performance of user activity detection (UAD) and 
channel estimation. The joint UAD and channel estimation problem 
can be formulated
and solved using compressive sensing or the Bayesian inference
approaches for better performance. 

\subsection{Machine Learning based GF Random Access}

Machine learning (ML) has been considered a powerful tool 
to solve complicated problems in a data-driven fashion. 
In GF random access, ML can be used 
for modeling overall processes without explicit 
mathematical formulations.
Based on this, optimization problems 
for improved throughput or reliability of GF random access 
schemes can be formulated and solved by applying machine learning, 
which overcomes the difficulties due to 
the mathematical intractability when using non-learning methods.

Obviously, the effectiveness of ML-aided GF random access 
makes it a promising solution to massive access in 6G networks. 
However, the diverse technologies and quality-of-service (QoS) 
requirements of applications that are envisioned for 
6G impose challenges for ML designs. In particular, the dynamic nature 
in the number of active devices 
and services across slots becomes more 
significant. 
Thus, an efficient 
and unified ML framework is needed for large-scale 
and high dynamic 6G GF random access systems. Motivated by the fact 
that most devices generally 
fall into some types of classes where they share similar properties, e.g., the same QoS requirements, efficient ML implementations can be attained via a multi-task learning approach. In this 
approach, each device class is regarded as a cluster 
and GF resource (GFR) 
can be dynamically partitioned into several sub-GFRs. 
All devices in each cluster will select resource within each sub-GFR.
Deep reinforcement learning (DRL) can be used to partition
and allocate
GFR adaptively. Also, the BS and 
devices can use ML methods to perform active device detection and 
GFR selection, respectively. This solution can promise to improve the training efficiency of the ML scheme, provide the flexibility of resource allocation, 
and reduce the complexity of data decoding.

To further improve the performance of ML-based 
GF random access schemes, temporal correlation characteristics 
of devices, i.e., sparse activities and sporadic traffic, should be exploited. 
Indeed, some classes of 
devices in practice are generally active 
in a regular mode with deterministic activity patterns. 
While the other classes of devices behave randomly, 
their transmit signals are usually correlated 
in several continuous time slots.
Hence, these characteristics can be learned and predicted to some extent. 
One of the promising solutions to implement this 
is using long short-term memory (LSTM)-based ML, 
a learning scheme that is able to remember a single 
event for long time periods. In particular, 
the history data about activities of MTC devices in previous 
transactions can be used 
to train by the BS for activity and traffic prediction in the current slot. 
Furthermore, devices with LSTM-based DRL algorithms 
can learn from their previous resource selections for 
efficient selections of preambles/sequences or power levels for 
GF random access. 
Finally, it would be important to take into account 
device activity and traffic patterns for 
efficient ML-based GF random access design.

\section{Conclusions}

GF random access
has been explained and its use with massive MIMO was discussed
in this article. While GF random access
can be more efficient 
in terms of spectral efficiency
due to a high spatial multiplexing gain with 
massive MIMO, its performance 
is limited by the number of preambles.
Thus, we explained a few approaches that can generate 
a large number of preambles.
Variations of GF transmissions
were also presented, 
which may allow us to design hybrid systems where
GF and grant-based schemes can co-exist.
A number 
of challenges of GF random access towards
next generation, i.e., 6G, were identified
and discussed.

\bibliographystyle{ieeetr}
\bibliography{mtc}

\end{document}